\documentclass[fleqn,10pt]{wlscirep}
\title{Coupling a single solid state quantum emitter to an array of resonant plasmonic antennas}

\author[1,2,3]{Markus Pfeiffer}
\author[4,5]{Paola Atkinson}
\author[4]{Armando Rastelli}
\author[4]{Oliver G.~Schmidt}
\author[3]{Harald Giessen}
\author[2,3,6,+]{Markus Lippitz}
\author[1,2,3,*]{Klas Lindfors}

\affil[1]{Department of Chemistry,  University of Cologne, Luxemburger Str. 116, D-50939 K\"{o}ln}
\affil[2]{Max Planck Institute for Solid State Research, Heisenbergstrasse 1, D-70569 Stuttgart}
\affil[3]{Fourth Physics Institute and Research Center SCOPE, University of Stuttgart, Pfaffenwaldring 57, D-70550 Stuttgart, Germany}
\affil[4]{Institute for Integrative Nanosciences, IFW Dresden, Helmholtzstrasse 20, D-01069 Dresden, Germany}
\affil[5]{Sorbonne Universites, UPMC Univ Paris 06, CNRS, UMR 7588, Institut des Nanosciences de Paris, 4 place Jussieu, F-75252 Paris, France}
\affil[6]{Experimental Physics III, University of Bayreuth, Universit\"atsstrasse 30, D-95447 Bayreuth, Germany}
\affil[+]{corresponding author, email: markus.lippitz@uni-bayreuth.de}
\affil[*]{corresponding author, email: klas.lindfors@uni-koeln.de}

\begin{abstract}
Plasmon resonant arrays or meta-surfaces shape both the incoming optical field and the local density of states for emission processes. They provide large regions of enhanced emission from emitters and greater design flexibility than single nanoantennas. This makes them of great interest for engineering optical absorption and emission. Here we  study  the coupling of a single quantum emitter, a self-assembled semiconductor quantum dot, to a plasmonic meta-surface. We investigate the influence of  the spectral properties of the nanoantennas and the position of the emitter in the unit cell of the structure. We observe a resonant enhancement due to emitter-array coupling in the far-field regime and find a clear difference from the interaction of an emitter with a single antenna.
\end{abstract}
\begin{document}

\flushbottom
\maketitle
%
%
\thispagestyle{empty}


\section*{Introduction}
Plasmon resonant nanoparticles enable propagating light fields to be converted into localized energy and \emph{vice versa}~\cite{Muehlschlegel2005,novotny:2011}. Due to this property they are called optical antennas. Optical antennas allow the impedance between a subwavelength light source, such as a semiconductor quantum dot (QD), and free space to be matched, resulting in increased emission~\cite{Farahani:2005,Kuehn2006,Anger2006,Taminiau2008,Li2010,Akselrod2014}. Optical antennas also enable the radiation pattern to be shaped~\cite{Curto:2010,Dregely:2014} and the polarization of the emission to be modified~\cite{Biagioni2009,Taminiau2007}, making them an interesting tool to control light-emission from single quantum emitters.

In order to significantly modify the spontaneous decay of a quantum emitter using optical antennas, the emitter has to be placed in the near-field of the plasmonic nanostructure. For quantum emitters buried in a substrate such as self-assembled semiconductor quantum dots or nitrogen-vacancy centers in a diamond crystal this requires the emitters to be placed very close to the substrate surface. This often results in degradation of the optical properties of the emitters due to surface states~\cite{Wang2004,Wang2016}. Additionally, the plasmonic nanostructure has to be positioned or fabricated with extremely high spatial accuracy with respect to the quantum emitter, which is often challenging~\cite{Badolato2005,Dousse2008,Schneider2008,Thon2009,Pfeiffer2014}. A promising alternative to optical antennas to control light emission from emitters that may resolve these challenges is \emph{optical metasurfaces}~\cite{Kildishev2013,Lin2014,Yu:2014}. Here plasmonic antennas are arranged in arrays~\cite{Felidj:2005,Liu:2008,Yang:2015} that allow controlling the spatial distribution of the amplitude, polarization, and phase of the electromagnetic field with subwavelength spatial resolution~\cite{Saito:2015}. In plasmonic nanoantenna arrays there exist large regions of enhanced emission and absorption in the unit cells of the array, so that position dependence of the coupling is less critical~\cite{Lozano2013,Rodriguez2012}. This could be highly desirable for plasmonic emission enhancement for organic optoelectronic devices and other thin film structures. In this study we investigate the coupling of single self-assembled semiconductor quantum dots to arrays of plasmon resonant nanoantennas. We observe modifications of the emission when the optical resonance of the array is tuned to the emission energy of the emitters, and demonstrate enhanced emission for quantum dots placed in the array outside of the near- and intermediate-zones of the plasmonic antennas.

\begin{figure}
\centering
\includegraphics[width=80mm]{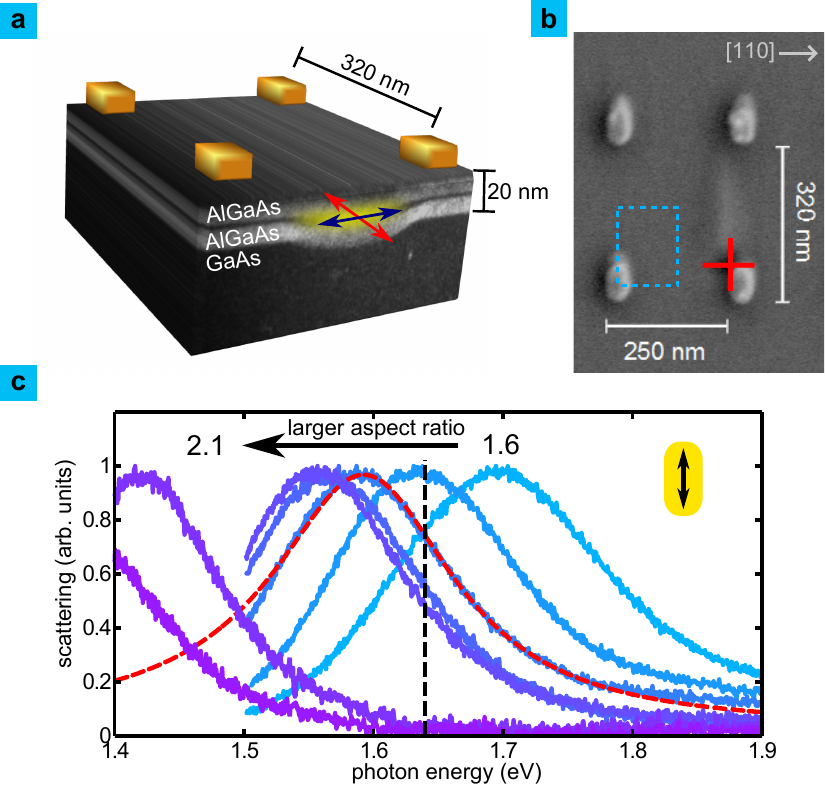}
\caption{Excitons in GaAs quantum dots couple to plasmons in a rectangular array of gold nanorods. (a) Arrays of rectangular nanoantennas are fabricated on near-surface GaAs quantum dots. The unit cells of the arrays are centered on the antenna elements. The orientations of the dipole moments associated with the two bright neutral exciton transitions are shown with blue and red arrows. (b) A characteristic topography feature is visible above the quantum dots in scanning electron micrographs (red cross). The x-axis corresponds to the [110] crystal direction. A quarter of the unit cell is indicated by the blue dashed frame. (c) We tune the longitudinal plasmon resonance of the array (solid lines) through the quantum dot exciton transition (black dashed line) by varying the nanorod aspect ratio. The nanoantenna scattering spectra are well represented by a Lorentzian fit (dashed red line).}\label{fig:Fig1}
\end{figure}

\section*{Results}
We prepare rectangular arrays of plasmonic gold nanorod antennas with varying aspect ratio on the surface of a crystal containing near-surface GaAs quantum dots (see Fig.~\ref{fig:Fig1}a). 
Each plasmonic nanoantenna array has an area of 100~$\mu$m~$\times~$100~$\mu$m, containing approximately $5\times 10^3$ quantum dots. The period of the array is 250~nm in the transverse direction of the nanorod antennas and 320~nm or 330~nm in the longitudinal direction (see Fig.~\ref{fig:Fig1}b). For the quantum dot emission energy (wavelength $\approx$~760~nm) these subwavelength periods result in only evanescent grating orders for the [1,0] and [0,1] array directions in the air half space. The smallest plasmonic antennas have in-plane dimensions of 50~nm and 82~nm (aspect ratio 1.6). We vary the aspect ratio of the nanorods to tune the longitudinal plasmon resonance of the array from 1.7~eV to below 1.4 eV as illustrated by the dark-field scattering spectra shown in Fig.~\ref{fig:Fig1}c. Here the incident s-polarized light has the electric field along the long axis of the antennas, so that only the longitudinal plasmon oscillation is excited. Full-field simulations using a finite-element method based solver (Comsol Multiphysics) show that this excitation is the fundamental dipolar mode. For the incident polarization along the transverse direction we observe no resonances close to the spectral position of the quantum dot emission. We finally note that the gold nanorods, and thus the arrays' [1,0] ([0,1]) directions are aligned with the [110] ([1-10]) substrate crystal directions, and thus with the transition dipoles of the quantum dots (Fig.~\ref{fig:Fig1}a). This has the advantage that the coupling of the two independent excitonic transition dipoles to the modes of the antenna arrays can be investigated.

\begin{figure}
\centering
\includegraphics[width=95mm]{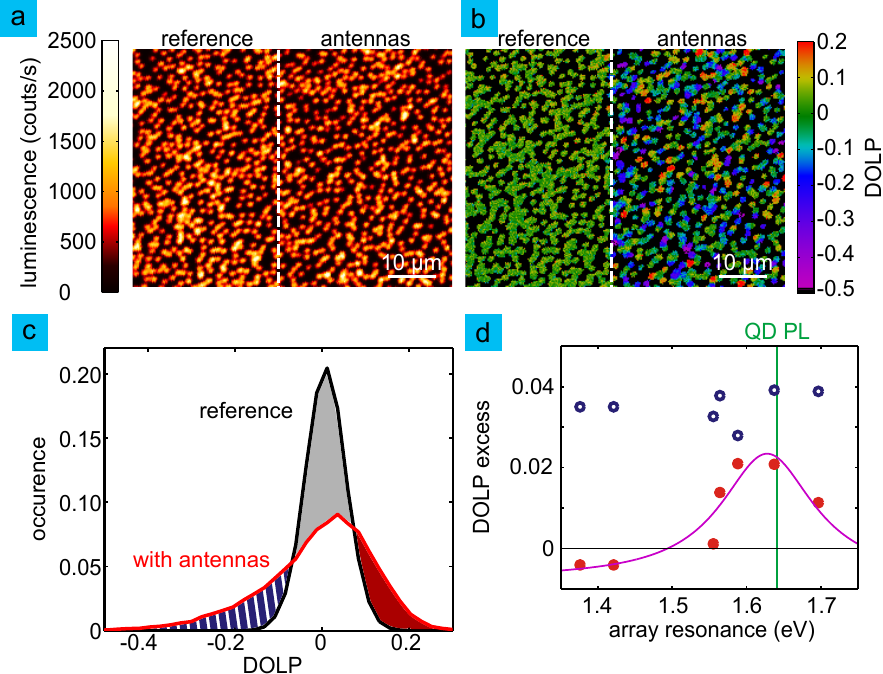}
\caption{Coupling to an array of optical antennas results in modified photoluminescence. (a) The total emitted photoluminescence shows only small differences between the sample areas with and without optical antennas. (b) For the same area the degree of linear polarization (DOLP) varies significantly due to the plasmonic nanostructures. (c) The histogram of DOLP values is significantly broadened for the sample area with optical antennas (red) as compared to the reference region (black). A fraction of the reference distribution (gray shading) is shifted to negative and positive values (blue and red shading, respectively).  The data in (b) and (c) is for the array with longitudinal plasmon resonance tuned to the emission of the quantum dots as evidenced by the scattering spectrum shown in Fig.~\ref{fig:Fig1}c (black dashed line indicates the quantum dot emission). (d) We quantify the influence of the resonance position in terms of  excess in the positive (red circles) or negative (blue circles) part of the DOLP histogram. The wavelength dependence of the positive excess follows a Lorentzian curve centered on the quantum dot emission while the negative excess is almost wavelength independent.}
\label{fig:Fig2}
\end{figure}

In photoluminescence images of quantum dot ensembles the emitted power shows only small differences between the sample area with and without optical antennas as seen in Fig.~\ref{fig:Fig2}a. We therefore turn to polarization resolved experiments and make use of the fact that the response of the plasmonic array is strongly polarization-dependent. We consider changes in the degree of linear polarization DOLP of the quantum dot luminescence, defined as
\begin{equation}
\mathrm{DOLP} = \frac{s_\mathrm{res}-s_\mathrm{nres}}{s_\mathrm{res}+s_\mathrm{nres}}.
\end{equation}
Here $s_\mathrm{res}$ and $s_\mathrm{nres}$ are the measured luminescence signals for emission polarized in the longitudinal (resonant) and transverse (non-resonant) direction, respectively. For off-resonant excitation at a photon energy of 2.327~eV, quantum dots without plasmonic structures display a narrow distribution of the DOLP centered at zero. The DOLP is furthermore not influenced by possible enhancement of the excitation and only reflects modifications of the emission. Figure~\ref{fig:Fig2}b shows the DOLP for the same region from where the intensity data ($s_{res} + s_{nres}$) of Fig.~\ref{fig:Fig2}a is acquired. We now observe significant changes in the emission due to the coupling to the plasmonic array that results in increased emission of the exciton with its transition dipole moment along the longitudinal direction.

To investigate the influence of the detuning of the plasmonic nanoantenna resonance with respect to the quantum dot emission we record photoluminescence images (integrated intensity from 1.570~eV to 1.656~eV) of the plasmonic arrays and include an area of the neighboring unpatterned region to act as reference in the images (right and left sides of \ref{fig:Fig2}a,b, respectively). We analyze the recorded images by first excluding pixels where the intensity is below a threshold value and which correspond to detector dark counts. We then calculate the DOLP for the remaining pixels and represent the data obtained in this way as (normalized) histograms. Figure~\ref{fig:Fig2}c shows the data for the array that is resonant with the exciton transition. Compared to the distribution of the reference area without plasmonic nanoantennas we observe a significant change in the shape of the histogram. The coupling between the quantum dot exciton and the plasmonic mode results in quantum dots that have a DOLP differing significantly from the reference emitters that display almost no polarization anisotropy~\cite{Pfeiffer2010,Pfeiffer2012}.

To quantify the change in the DOLP distributions we calculate, for arrays with different resonance energies, the excess positive and negative values of the DOLP compared to a reference sample as first moments of the positive and negative parts of the DOLP distribution (see supporting information for more detailed information). Figure~\ref{fig:Fig2}d shows the dependence of these excess values on the detuning of the nanoantenna array resonance from the quantum dot emission. We note that the excess positive values of DOLP closely follow the resonance of the plasmonic structure while the negative values are almost independent of the properties of the nanoantenna array. We next interpret these results based on simulations and position-resolved measurements of the DOLP.

\begin{figure}
\centering
\includegraphics[width=120mm]{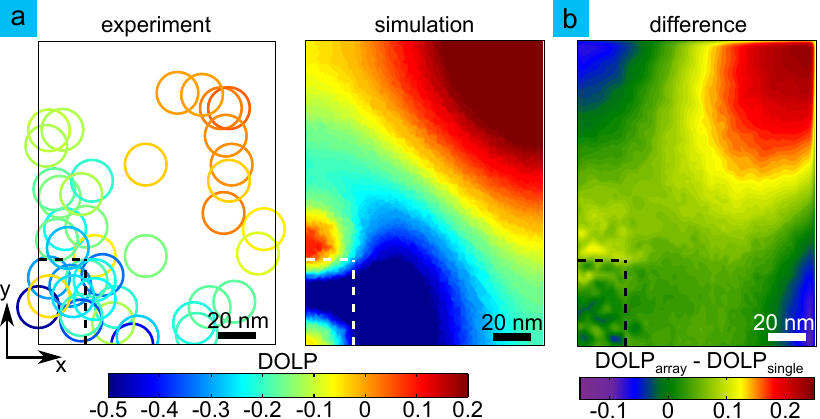}
\caption{The DOLP shows a distinct distribution in the unit cell of the plasmonic nanoantenna array. Here the dashed rectangle signifies a plasmonic nanoantenna. (a) The measured DOLP (color coded circles) for the quantum dot emission (left) is negative near the antenna and along the longitudinal direction where the near- and intermediate-fields are strong. For larger separations, where far-field coupling is significant, the emission is enhanced, resulting in elevated DOLP values. Finite element simulations for the plane of the quantum dots (right) are in good agreement with the experimental data. Here we consider a quarter of the unit cell of an array of plasmonic antennas with longitudinal modes resonant to the quantum dot emission (see Fig.~\ref{fig:Fig1}). (b) The difference in the calculated DOLP distributions for an array and a single antenna is small near the antenna and becomes significant further away. The positive DOLP in the corners of the unit cell can be clearly attributed to interactions of array elements.}\label{fig:Fig3}
\end{figure}

\subsection*{Individual quantum dots in resonant array: Position-resolved interaction}

Let us now turn to the position-dependent coupling of a single quantum dot to the plasmonic array. We use the characteristic topography feature above each quantum dot to determine the position of quantum dots with respect to the plasmonic array~\cite{Pfeiffer2014}. We show color-coded DOLP values for 43 quantum dots as a function of position on the left side of Fig.~\ref{fig:Fig3}a. The colored rings represent experimental data for different quantum dot positions. As the optical properties of the self-assembled quantum dots and the optical antennas exhibit narrow ensemble distributions, the results for all structures are visualized in one plot. The radius of the ring corresponds to 11~nm, which is the worst case estimate for the accuracy with which we can determine the position (see Ref.~16). Here the quantum dot positions are plotted in a quadrant of the unit cell of the array. The position-resolved measurements point to the coupling mechanism: Beside the antenna we observe a suppression of emission for the resonant polarization signal  of the y-dipole due to quenching together with an increase of emission for the transverse polarization (x-dipole) and therefore low values of DOLP. The small enhancement of the x-dipole is likely due to coupling to the transverse plasmon, which however is significantly detuned from the quantum dot emission. As both effects depend only weakly on the element length (longitudinal resonance) the excess negative part of the DOLP histogram is almost independent of the spectral position of the array resonance (see Fig.~\ref{fig:Fig2}d). The positive DOLP originates from enhanced emission from quantum dots located in the corners of the unit cell furthest away from the antennas. We compare the position-resolved experimental data to simulations. The simulated data is obtained by using the Lorentz reciprocity theorem~\cite{Carminati1998}. The theorem relates the electric fields radiated by dipoles placed at the position of the quantum dot and detector at these locations. We can in this way deduce the detected intensity of a dipole emitter as a function of its relative position to a plasmonic structure~\cite{Pfeiffer2014}. We obtain the DOLP by taking the two orthogonal polarizations of the quantum dot transition dipoles into account. The right plot in Fig.~\ref{fig:Fig3}a shows the simulated DOLP distribution. We observe good agreement between the experiment and theory. More insight is gained by looking at the difference in the DOLP distribution for a single antenna and an array as shown in Figure~\ref{fig:Fig3}b. We find that the differences in the distributions are insignificant for near-field coupling while there are significant differences for larger separations between antenna and emitter. In the diagonal [1,1] direction at a distance more than 50~nm away from the antenna intermediate- and far-field coupling is dominant and we obtain enhanced emission of the polarization parallel to the resonant array direction of up to a factor of approximately two (data not shown here). For the intermediate-field coupling along the resonant [0,1] array direction the emission is suppressed. See also Fig. S3 in supplementary information.



\section*{Discussion}
In summary, we have studied the coupling of single semiconductor quantum dots to an array of rod-shaped plasmonic nanoantennas. We observe a polarization anisotropy of the emission that is dependent on the spectral position of the plasmon resonance with respect to the narrow-band quantum dot emission. The coupling-induced polarization change can be separated into wavelength-independent and resonant components: The former is a decrease of the emission polarized along the plasmon oscillation for quantum dots within the near- and intermediate-zones of the antenna. The latter is a resonant increase of the emission for quantum dots further away from the plasmonic rod. Using full-wave simulations we demonstrate that the resonantly enhanced emission originates from coupling of the exciton in the quantum dot to plasmons in the \emph{array}. For a single antenna the long-range enhancement is lacking.

We show a strongly position and resonance dependent polarization anisotropy which originates from the interaction in different field regimes. Our work demonstrates a novel approach to enhance emission from self-assembled semiconductor quantum dots using plasmonics. Quenching of emission present in near-field coupling is avoided when using plasmonic nanoantenna arrays. This enables increasing the emission and controlling the polarization and radiation pattern using suitably designed nanoantenna \emph{metasurfaces}. By making use of dielectric metasurfaces the large absorption present in plasmonic structures can be avoided~\cite{Kildishev2013,Lin2014,Yu:2014}. Our work shows that nano-antenna arrays are an alternative to single optical antennas to enhance and control the emission of single quantum emitters with the potential of greater flexibility in design and possibly lower losses.

\section*{Methods}

\subsection*{Growth of GaAs quantum dots}
These emitters are epitaxially grown using molecular beam epitaxy and are located at a depth of 21~nm inside the semiconductor crystal enabling coupling to plasmonic structures fabricated on the sample surface. The strain-free quantum dots~\cite{Atkinson:2012} display neutral exciton emission with a photon energy of approximately 1.64~eV and their almost degenerate transition dipoles are aligned along the [110] and [-110] crystal directions. The growth and optical properties have been reported in detail in Refs.~\cite{Atkinson:2012,Pfeiffer2010,Pfeiffer2012}. 

\subsection*{Fabrication of plasmonic nanoantenna arrays}
The plasmonic nanoantennas are fabricated on the sample surface using electron beam lithography. We first spin coat a 200 nm thick double layer PMMA (polymethyl methacrylate) resist where the bottom layer is more sensitive than the 50~nm thick top layer. This results in an undercut of the resist after development, assisting lift-off. The resist is exposed with 20~kV electrons and developed followed by metal deposition by thermal evaporation. Here we use a 3~nm thick chromium layer for improved adhesion followed by 30~nm gold. A lift-off process removes the excess metal. We have previously shown that the optical properties of the quantum dots are not modified in this process~\cite{Pfeiffer2014}.

\subsection*{Optical measurements}
The photoluminescence was measured in a home-built low-temperature laser-scanning confocal microscope. The samples were cooled down to 10-15~K in a liquid helium flow-cryostat. The numerical aperture of the objective used in the experiments was 0.7 and the polarization of the incident light is set to the transverse direction of the rod-shaped antennas. A full sketch of the setup is shown in Fig. S1 of the supplementary information.

\subsection*{Finite element simulations}
The simulations were carried out using a finite element solver (Comsol Multiphysics). Full details about the simulations are given in the supplementary information.


\section*{Acknowledgements }
We gratefully acknowledge funding from the DFG (research group FOR730) and the Academy of Finland (project 252421). This work was partly funded through the Institutional Strategy of the University of Cologne within the German Excellence Initiative. We furthermore would like to thank J.~Weiss and the Nanostructuring Lab team of the Max Planck Institute for Solid State Research for help with sample fabrication.

\section*{Author contributions statement}
M.P., K.L., M.L., and H.G. conceived the experiments,  P.A. grew the quantum dots samples under supervision of A.R. and O.S.,  M.P. and K.L. conducted the experiments, M.P., K.L., and M.L. analyzed the results. M.P. and K.L. carried out the FEM simulations. M.P. and K.L. wrote the manuscript with input from all authors.

\section*{Additional information}

\textbf{Competing interests:} The authors declare that they have no competing interests.

\noindent\textbf{Data availability statement:} The datasets generated during and analysed during the current study are available from the corresponding author on reasonable request.


\newpage
\newcommand*\mycommand[1]{\texttt{\emph{#1}}}

\renewcommand{\thefigure}{S\arabic{figure}}

\section*{Supplementary information: Coupling a single solid state quantum emitter to an array of resonant plasmonic antennas}


\section{Experimental setup}
\begin{figure}[b!]
\centering
\includegraphics{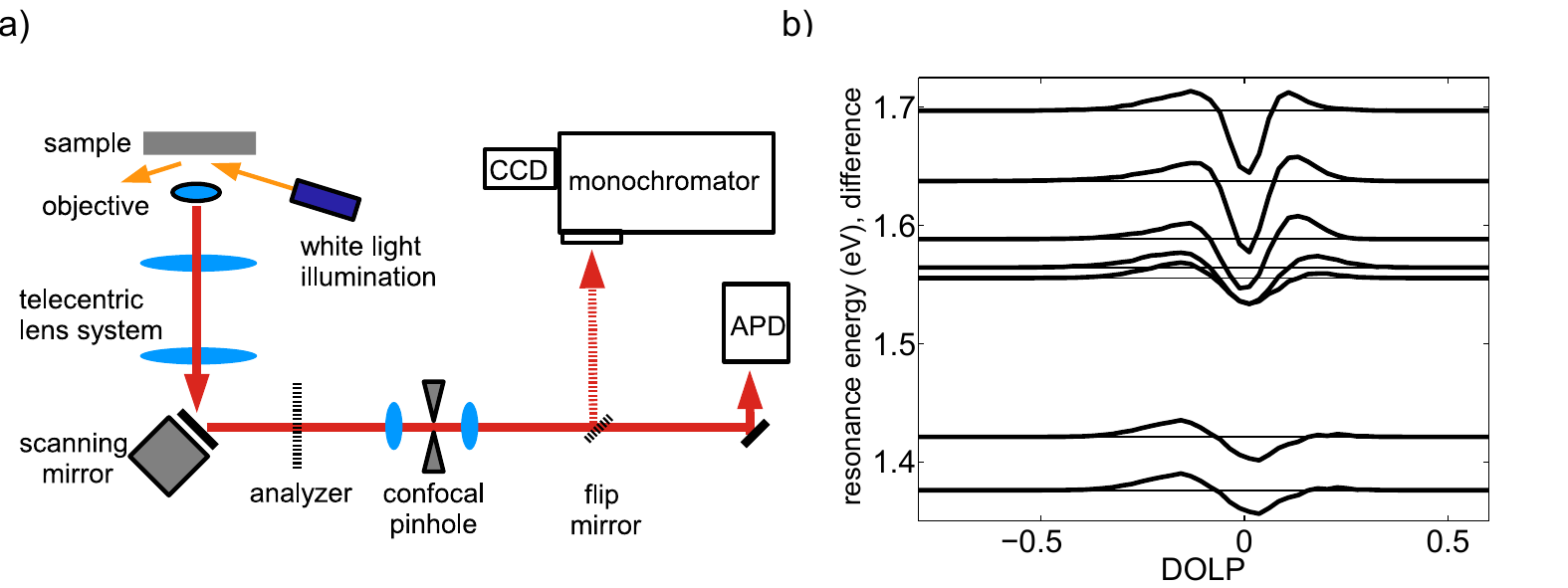}
\caption{a) Experimental setup. b) Difference of histograms for reference emitters and emitters beneath arrays with the indicated resonance energy. }\label{fig:FigS1}
\end{figure}

All optical measurements are performed in a home-built laser-scanning confocal microscope. This setup can also be used to perform detection-spot scanning dark-field micro-spectroscopy, as illustrated in Fig.~\ref{fig:FigS1} a). For plasmonic antennas on semiconductor substrates it is necessary to have control over the polarization state of the illumination to be able to distinguish between the different modes. To achieve this, a fiber-coupled illumination module was designed and implemented in a home-built laser-scanning confocal microscope. The illumination source is a 100~W halogen lamp that is coupled into a large core multi-mode fiber using a hemispherical lens with a focal distance of 20~mm. The light exiting the fiber is collimated, polarized with a broadband polarizer, and then slightly focused on the sample under an adjustable angle. Light scattered by an optical antenna is collected by the objective in the laser-scanning confocal microscope, passed through the confocal pinhole, and then directed either onto a single photon counting avalanche photodiode (APD) module or to a spectrometer equipped with a charge coupled device camera (CCD). The sample area from which light is detected can be controlled with the scanning mirror in the microscope.

\section{Details on quantifying the excess DOLP}
To quantify the change in the DOLP distributions we calculate the difference of the normalized histogram for the antenna array $N_\mathrm{res}$ and the unpatterned quantum dot region $N_\mathrm{ref}$. The photoluminescence signals of the quantum dots for both polarizations are spectrally integrated intensities from 1.570~eV to 1.656~eV. The differences of the histograms for all antenna arrays are displayed in Fig.~\ref{fig:FigS1}b). To quantify the changes due to the arrays we calculate the centers of mass for the differences of the histograms. We perform this analysis separately for positive and negative values of DOLP. This allows us to observe changes in the shapes of the histograms in order to gain insight into how the enhancement or suppression of emission depends on the properties of the plasmonic array. For positive values of DOLP we define the DOLP excess as 

\begin{equation}
m^+ =\\ \int_0^1 \left( N_\mathrm{res} - N_\mathrm{ref}\right)\mathrm{DOLP}~\mathrm{d(DOLP)}.
\end{equation}
For negative values of DOLP we define
\begin{equation}
m^- =\\ \;  -\int_{-1}^0 \left( N_\mathrm{res} - N_\mathrm{ref}\right)\mathrm{DOLP}~\mathrm{d(DOLP)}.
\end{equation}

\begin{figure}[t!]
\centering
\includegraphics{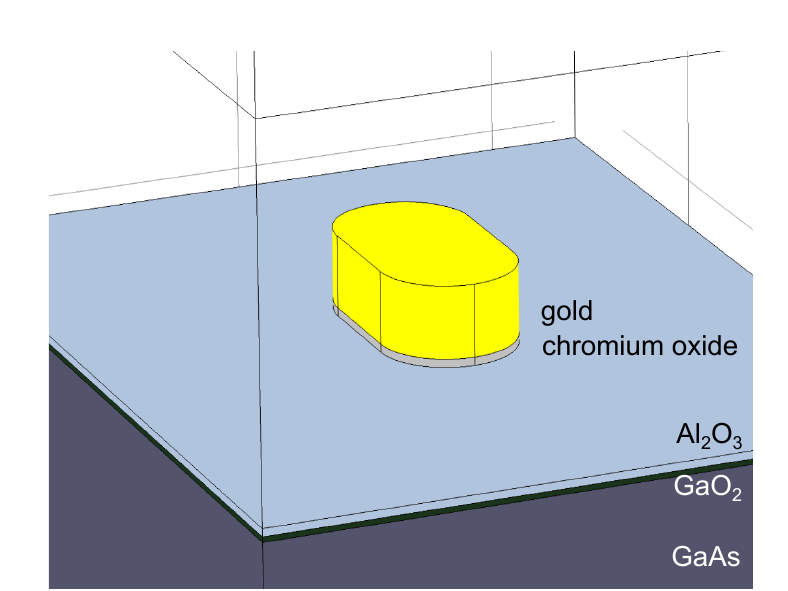}
\caption{Structure used for simulations}\label{fig:FigS2}
\end{figure}

\section{Finite element simulations}
Simulations are performed with a finite element solver (Comsol Multiphysics with RF module, Comsol Ab, Stockholm, Sweden)  for frequency domain electromagnetic full field calculations. As described in previous publications~\cite{Pfeiffer2014}, we apply the Lorentz reciprocity theorem to deduce the emitted power from oriented electric point dipoles~\cite{carminati:1998} at the position of the quantum dots. The periodicity is included by applying Floquet periodic boundary conditions on the side boundaries of the arrays' unit cells. 
For the calculations of the intensity distribution around an isolated single antenna, we apply a two step simulation. First, the incident fields are calculated, while the material of the antenna is set to vacuum. In a second step, the field of the first simulation step is taken as the incident field, the antenna is modeled as a nanorod of gold and chromium oxide, and periodic boundary conditions are replaced by perfectly matched layer (PML) domains with outer scattering boundary conditions.

The structure is modeled according to dimensions determined from scanning electron micrographs and atomic force microscopy measurements. The antennas are rectangular gold nanorods of 90~nm length, 60~nm width, and 27~nm height. The dielectric constants of gold are interpolated from Ref.~\cite{johnson:1972}. 
Beneath the antennas there is a 3~nm thick layer of chromium oxide (refractive index = 2.5). The substrate consists of an infinite substrate halfspace with a refractive index of 3.5 on top of which there is a 2~nm thick native gallium oxide layer (refractive index = 1.9)~\cite{Rebien2002}. At the sample surface the 3~nm thick Al$_2$O$_3$ passivation layer is modeled as a layer with a refractive index of 1.76. 
To adapt the element shapes to those determined for similarly fabricated gold structures from transmission electron micrographs~\cite{Pfeiffer2014}, the side corners of the antenna elements are rounded with a radius of 25~nm (see Fig.~\ref{fig:FigS2}).
From transmission electron micrographs we determine a distance of 21~nm between the center of the quantum dots and the top of the substrate. Thus, in the simulations we extract the intensity distributions for two orthogonal polarization states at this distance from the substrate air interface.

\begin{figure}[t!]
\centering
\includegraphics{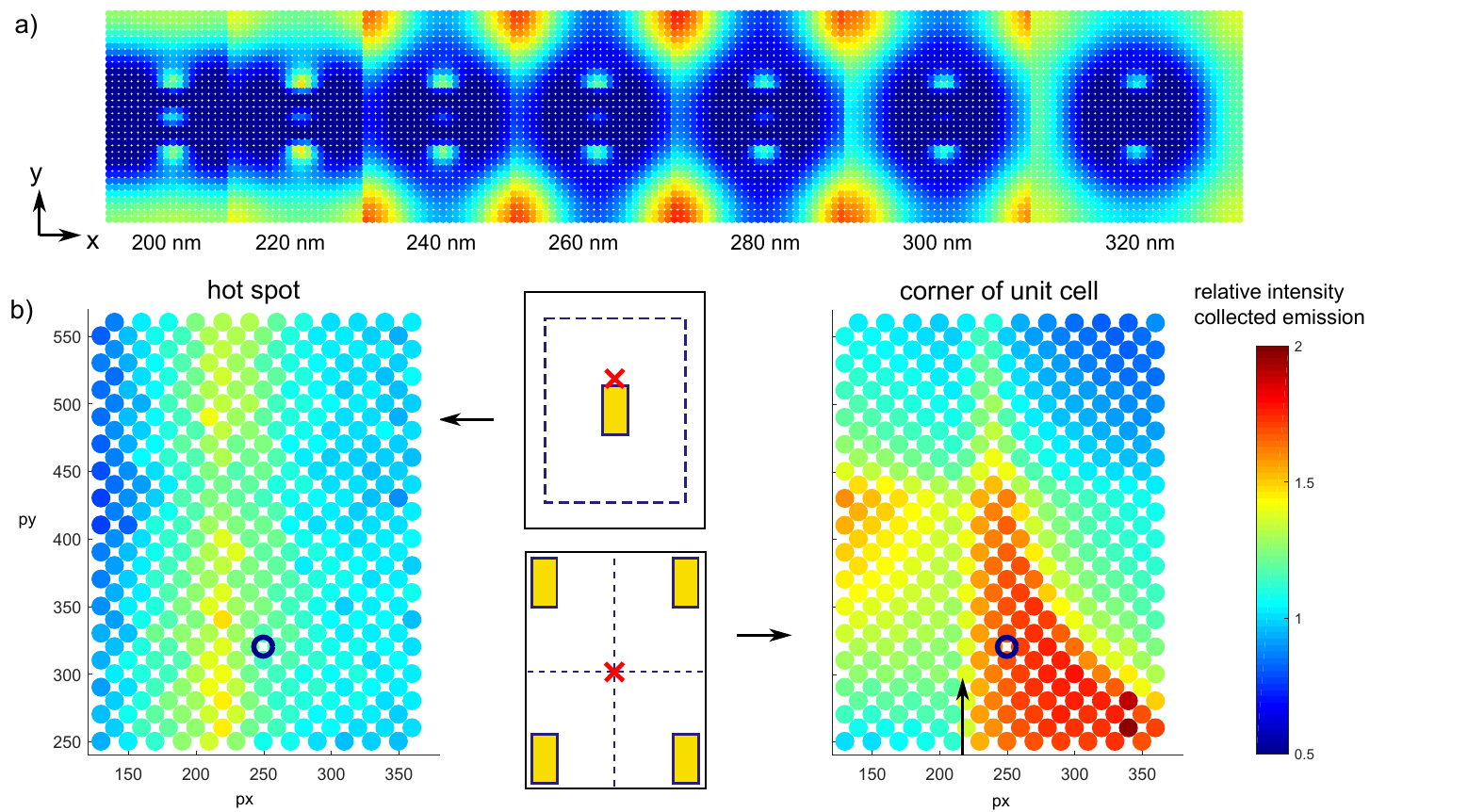}
\caption{a) Intensity distribution for a $y$-oriented dipole in the plane of the quantum emitters. The period in $y$-direction is kept constant (320~nm) while the period in $x$-direction is varied from 200~nm to 320~nm. b) Enhancement of detected emission for different array periods for a dipole located at a distance of 10~nm from the tip of the antenna elements (left) and in the corner of the unit cell (right). The vertical arrow marks the array period $p_x$ where the first diffraction order becomes radiative into the substrate. The arrays studied experimentally in this work are marked with a blue circle.}\label{fig:FigS3}
\end{figure}

\section{Influence of the array period}

Here we analyze, how the observed mode is influenced by the period of the plasmonic array. For this, we keep the emission wavelength constant at the QD emission ($\lambda_{0,QD}$ = 760~nm) and investigate the collected intensity of a $y$-oriented dipole, which is located in the plane of the quantum dots. For a constant period in $y$-direction (chosen identical to the experimental conditions, $p_y = 320$~nm) the intensity of the $y$-oriented dipole is shown as a function of its lateral position in the unit cell [see Fig.~\ref{fig:FigS3}a)]. One observes, that there are regions of enhanced emission close to the ends of the antenna elements and in the corners of the unit cell. For these two locations we extract the relative enhancement compared to an emitter in the same environment but without the plasmonic array. This is shown in Fig.~\ref{fig:FigS3}b). The enhancement for the quantum placed in the corner of the unit cell [see right side of Fig.~\ref{fig:FigS3}b)] decreases for increasing period in $x$-direction $p_x$. For a period $p_x$ of approximately 220~nm one can observe a discontinuity in the enhancement. This is due to the first diffraction order into the substrate becoming radiative as the condition $p_x > \lambda_{0,QD} / n_{substrate}$ is fulfilled. 

The deviations between the spatially dependent enhancement of experiment and simulations are mainly the lack of enhancement for the y-oriented dipole at the hot-spots of the antenna elements. We attribute this to quenching of the emission due to an additional absorption rate due to the proximity of the metal structure. This additional rate is not considered in our simulations. Another reason are geometrical mismatch between the size of the hot-spots and the lateral dimension of our quantum dots. The spatial averaging of the enhancement pattern results in a decrease of the calculated maximum values, where we considered point dipoles. Particularly, from Fig. S3 it becomes obvious, that the enhancement shows a different dependence on the lateral array periods than the hot-spots close to the long ends of the array elements.


\bibliography{One_against_many}

\end{document}